# Highlighting Impact and the Impact of Highlighting: PRB Editors' Suggestions

With worldwide R&D investment topping $1.5 trillion annually [1] and constantly increasing technical capabilities, computing power, and means of scientific interactions, global research output is growing rapidly (in fact, exponentially [2]) in numbers and scope. Just over the last 15 years, the number of scientific articles published annually in physics journals [3] rose from approximately 78,000 to 133,000. APS journals, in particular, now publish almost 18,000 articles per year, of which approximately 5,000 appear in PRB.

In this environment, scientific publishers face new challenges. To stay relevant and competitive amidst the deluge of information, publishers need to maintain a broad scope by expanding their portfolios to new and emerging fields, publish sufficient numbers of papers to capture market share, attract some of the *key* developments in each field, provide tools to assist readers in navigating efficiently to papers of interest and relevance, and reward authors of excellent papers by providing the appropriate visibility and publicity. While juggling such challenges, publishers need to remain mindful of their overarching responsibility, namely to maintain high publication standards through rigorous peer review.

Some publishers have responded to the new challenges by simply allowing their journals to grow. Others have launched new titles, from large-sized and interdisciplinary journals, to highly exclusive titles, to numerous niche journals of low or moderate impact catering to narrow subfields. For example, from 1999 to 2014, the number of physics journals [3] has increased from 253 to 389. (The APS, in particular, has launched four new journals in this period.)

Many publishers have also started providing *select sets* of papers (highlights) that are deemed to be of higher quality than the average paper in their source journals to accord them higher visibility (with the associated online search features). Sometimes, highlighted papers are accompanied by editors' summaries or experts' commentaries that elucidate their importance. It is indicative of the evolving landscape of scientific publishing that this practice of highlighting has proliferated especially in the last 15 years, and most of the major physics publishers have adopted it by now.

Starting in 2008 (a year after PRL initiated the idea) [4], PRB editors have been highlighting a few papers each week, which they mark as Editors' Suggestions. These papers, when published, feature prominently on our website, with an image and short introduction written by the editors, and are marked with a special icon in the Tables of Contents and in online searches. Papers are selected primarily on the basis of their importance and impact in their respective fields, or because they are deemed interesting or elegant, and are assessed by the editors, who utilize the referees' reports and their own understanding. More often than not, the highlighted papers receive strong endorsement from our referees in the review process. We typically highlight 5 papers out of 90 each week.

How do Editors' Suggestions fare in terms of citation impact? The standard disclaimers are warranted here. First, the 'true' impact of a paper goes beyond citation impact. Second, we do not highlight papers with the sole criterion of citability in mind—and even if we did, citations to specific papers are hard to predict anyway, as experience has shown. Third, it is difficult to assess the citation impact of an *individual* paper before the lapse of a considerable amount of time. Fourth, judging importance and impact always contains a subjective element. Fifth, citation practices depend on the "hotness" of a field or the size of a scientific community; some papers take a long time to garner citations, etc. While there is value to all these concerns, we are not the only ones to have witnessed [5] that sufficiently large *groups* of papers that perform well, in terms of citations, *in the first couple of years*, tend to attract considerable interest in their respective communities in the longer run.

Figure 1 shows that on average Editors' Suggestions receive significantly more citations than the other PRB articles. For example, of the papers published in 2012–2013, Suggestions received on average 5.9 citations in 2014, compared with 4.8 citations for Rapid Communications, and 3.7 citations for PRB as a whole. (The latter number is the PRB Impact Factor.) This stratification of citation impact (PRB Editors' Suggestions > PRB Rapid Communications > PRB journal) is consistent throughout all years. (In the figure, we include the year 2009 for completeness, even though its sample size is much smaller than for the following years, which makes its average citation metric more sensitive to a few highly cited papers. However, the median citation metric for 2009 is no different than for the following years, as we explain below.)

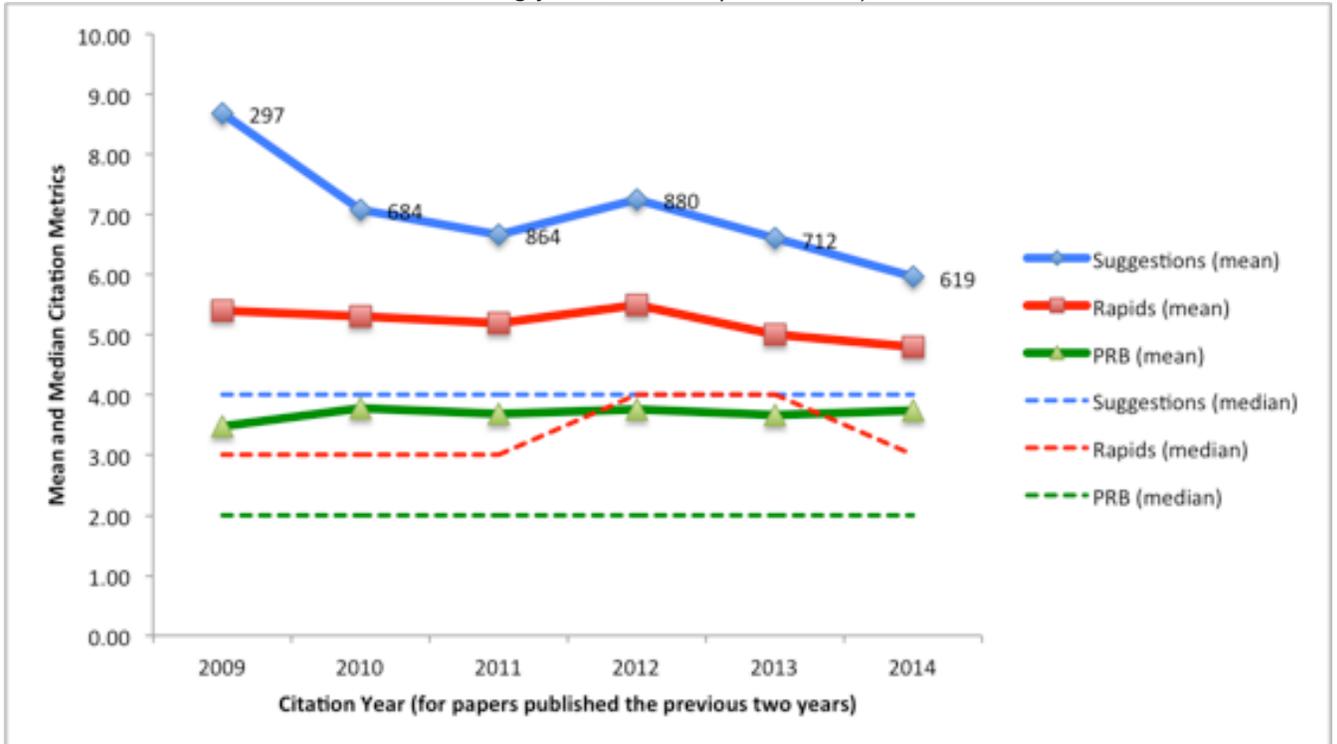

FIG. 1 (color online). Mean (solid lines) and median (dashed lines) citation metrics for PRB Editors' Suggestions (blue), PRB Rapid Communications (red), and the whole PRB journal (green). Citations are counted during the citation year shown on the horizontal axis, for papers published the previous two years. The mean citation metric for the whole journal is PRB's Impact Factor. The data labels on the blue line show the numbers of Suggestions published the two years prior to the citation year. Citation data from Web of Science. The measures of spread for Editors' Suggestions are as follows. Mean: The standard deviation ranges from 17.1 (2009) to 7.9 (2014). Median: The interquartile range (distance between 75th and 25th percentile) ranges from 5 to 7.

Is this elevated citation performance of Editors' Suggestions due to just a few outliers? To answer this question we look at median, as opposed to mean, citation counts. As shown in Fig. 1, median citations of PRB Suggestions are consistently twice those of the whole PRB journal in the time interval of choice. This result helps us see beyond annual fluctuations in the average citation metric of Editors' Suggestions (which approximately follow the fluctuations of the top-cited portion of the journal) and recognize the fact that the *typical* Suggestion gets cited well above the typical PRB article.

Going beyond means and medians, it is important to realize that there is a subset of very well cited papers among Editors' Suggestions. For example, of the 457 PRB papers (published in 2008–2014) that are classified by Thomson Reuters as ESI Top Papers [6], 133 were Suggestions. Thus, each PRB Suggestion is six times more likely to become an ESI Top Paper than a non-Suggested paper (5.5% of PRB Suggestions are ESI top papers, compared to 0.9% of the rest of PRB papers). Looking at the more recent batch, there were more than 100 Editors' Suggestions published in 2012–2013 that were cited 10 times or more in 2014; in terms of citability, these papers rank at or above the 93rd percentile of the whole PRB journal.

Of course, there are many papers that are highly cited and, indeed, important to their fields, that are not highlighted, just as there are some highlighted papers that are little cited. This feature, which is actually common, to varying degree, in all selection processes, should not distract us from the fact that, as shown from the numbers above, the selection process of Editors' Suggestions identifies many papers that will turn out to be influential.

How has the community responded to our highlighted papers? We often receive enthusiastic comments from authors whose papers we have selected, or explicit recommendations from referees endorsing papers for highlighting. We feel that now is a good time to go beyond anecdotal endorsements and make the case to our wider readership that Editors' Suggestions stand one notch above the rest of the journal. In an era where science is characterized by deep specialization but also wide interdisciplinarity, and where scientists face increasing competition for recognition while publishers compete for readers' time, broad scope journals such as PRB, through highlighting papers, can bring the best of each field to the attention of the wider readership. As the production of research publications continues its inexorable growth, we are hopeful that the PRB community will further recognize the additional layer of scrutiny—and the elevated citation impact—involved in the selection of papers as Editors' Suggestions.

Manolis Antonoyiannakis
Associate Editor